\documentclass[twocolumn,12pt]{article}

\usepackage{jeosman}
\usepackage{geometry}
\usepackage{graphicx}
\usepackage{color}
\usepackage{cite}

\begin{document}

\twocolumn[ 

\title{Incoherent interaction of nematicons in bias-free liquid-crystal cells}

\author{Yana Izdebskaya, Vlad Shvedov, Anton Desyatnikov, and Yuri Kivshar}

\address{Nonlinear Physics Center, Research School of Physics and Engineering, The Australian National University, Canberra ACT 0200, Australia}

\author{Wieslaw Krolikowski}

\address{Laser Physics Center, Research School of Physics and Engineering, The Australian National University, Canberra ACT 0200, Australia}

\author{Gaetano Assanto}

\address{NooEL, Department of Electronic Engineering, University of Rome ``Roma Tre'', 00146 Rome, Italy}

We study experimentally the propagation dynamics and interaction of a pair of mutually incoherent nematicons: spatial optical solitons in nematic liquid crystals. In contrast to earlier studies, we consider a bias-free liquid-crystal cell and compare the soliton interaction in copropagating and counterpropagating geometries. We analyze the dependence of nematicon interaction on input power and observe a direct manifestation of a long-range nonlocal nonlinearity. Attraction of counterpropagating solitons requires higher powers and longer relaxation times than that of copropagating nematicons due to losses-induced power asymmetry of counterpropagating nematicons.

\vspace{5mm}

]

\section{Introduction}

Solitons have been observed in diverse fields of nonlinear physics, and they share common fundamental properties originating from the interplay between nonlinear self-action of wave packets and their natural tendency to spread as they propagate. Spatial optical solitons~\cite{Kivshar}, i.e. nonspreading self-localized beams with the width unchanged during propagation, form due to a balance between linear diffraction and self-focusing in a nonlinear optical medium. These solitons have been investigated extensively in several nonlinear media, both in one- and two-dimensional geometries; they have significant potentials in signal processing, switching and readdressing in the future generation of all-optical circuits. In this context, a giant optical nonlinearity arising from molecular reorientation in nematic liquid crystals (NLCs) has attracted significant attention~\cite{Tabirian, Khoo,Simoni}. Both experimental~\cite{Peccianti2000, Hutsebaut} and theoretical~\cite{McLaughlin1996} results have been demonstrated for spatial optical solitons in nematics, also called {\em nematicons}~\cite{Assanto_OPN}.

Nematic liquid crystals consist of elongated molecules aligned along a given direction (known as the molecular director) owing to both anchoring at the boundaries and intermolecular forces~\cite{Tabirian, Khoo, Simoni}. The resultant medium exhibits a positive uniaxial anisotropy and birefringence, with ordinary and extraordinary refractive indices, $n_\parallel$ and $n_\perp$, defined for polarizations parallel and orthogonal to the director. The reorientational nonlinearity allows generating nematicons at relatively low optical powers, in the milliWatt region or below, for the study of the fundamental aspects and applications of light interaction with self-assembling nonlinear soft matter.

While initial studies of solitons in nematic liquid crystals considered the propagation of single nematicons~\cite{Peccianti2000, Peccianti2001, Conti2003, Assanto2003, Conti2004, Peccianti2004, Alberucci2005}, the interaction of two copropagating nematicons of equal~\cite{Peccianti2002,Fratalocchi2004,Fratalocchi2007,Fratalocchi2007PRE,Smyth2008} or different wavelengths~\cite{Alberucci2006, Assanto2008} have been addressed more recently. The attraction and fusion of nematicons in bulk undoped NLCs in planar voltage-biased liquid crystal cell were reported by Peccianti {\it et al.}~\cite{Peccianti2002, Peccianti2002_APL}.

\begin{figure}
\centering\includegraphics[width=8cm]{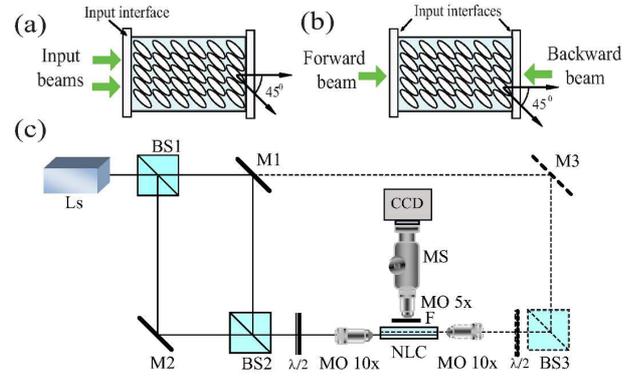}
\caption{Top view of the NLC planar cell for the study of (a) copropagating and (b) counterpropagating solitary beams. The ellipses indicate the orientation of the molecular director in the plane $(y,z)$. (c) Experimental setups for the copropagating (solid lines) and counterpropagating (dashed lines) cases: Ls -- cw-laser ($\lambda=532nm$), BS -- beam splitters, M -- mirrors, $\lambda/2$ -- half wavelength plates, MO -- microscope objectives, NLC -- cell, F -- filter, MS -- microscope, CCD -- camera.}
\label{fig1}
\end{figure}

Most of the studies considered nematicons propagating in the same direction. At the same time, the interaction of counterpropagating solitons in, e.g., photorefractive crystals~\cite{MotiPRL,Moti02,CP0,CPan}, is known to posses rich physical behavior, including convective dynamical instabilities~\cite{CP1,Moti04,CP2,DenzOE04}. The latter were observed in experiment in Ref.~\cite{DenzPRL} and stabilization against these instabilities was achieved in transversely periodic nonlinear medium~\cite{DenzOE07}. In contrast, there is little known about counterpropagating nematicons, although an experimental approach was developed, based on the mutual deflection of two counterpropagating beams, which permitted to estimate the strength of thermal focusing in thick dye doped nematic-liquid-crystal samples excited by narrow laser beams~\cite{Henninot2002, Henninot2007}. In addition, recent numerical studies pointed out the possible occurrence of instabilities for counterpropagating beams in liquid crystals~\cite{Strinic2006, Strinic2008}.

\begin{figure}
\centering\includegraphics[width=7.95cm]{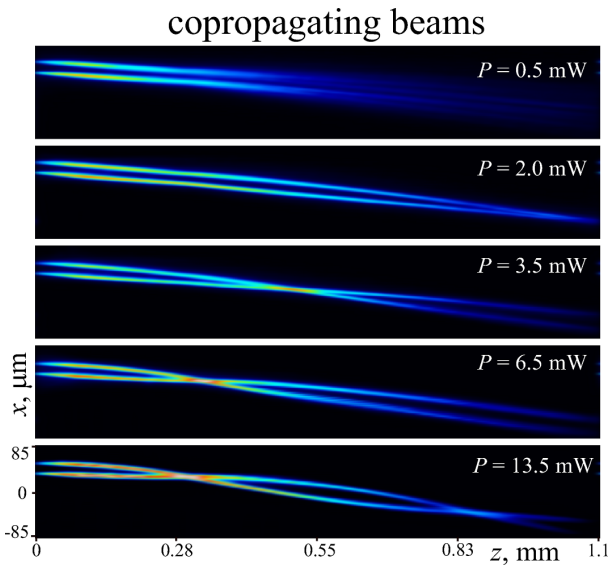}
\caption{{\color{blue}} Experimentally recorded images of light scattered from two CO nematicons at various input powers $P$ in each beam.}
\label{fig2}
\end{figure}

In earlier experimental studies with nematicons the liquid-crystal cell was biased by an external electric field, aimed at controlling the orientation of the NLC molecules at rest and, therefore, their nonlinearity as well as their nonlocality~\cite{interplay}, thus also affecting nematicon interactions. However, nematicons can exist in unbiased NLC cells provided the input beam is extraordinary polarized, e.g. with director and electric field coplanar with the plane $(x, z)$ parallel to the cell interfaces~\cite{Peccianti2004, Alberucci2005, Alberucci2006, Fratalocchi2007, Fratalocchi2007PRE, Piccardi2008}. Because of the optical anisotropy of liquid crystalline molecules and birefringence, light beams propagating in NLC walk-off the direction of their wave vector. The walk-off can be adjusted by acting on the optic axis, i.e. by reorienting the molecular director ~\cite{Peccianti2004,Peccianti2005}.

In this paper we experimentally address the propagation and interaction of copropagating (CO) and counterpropagating (CP) nematicons in a bias-free NLC cell. We investigate in detail both power- and time-dependent interactions of two identical solitons. In particular, we demonstrate that the long-range attraction and ability of nonlocal solitons to form bound states are strongly affected by the geometry of the interaction.

\begin{figure}
\centering\includegraphics[width=8cm]{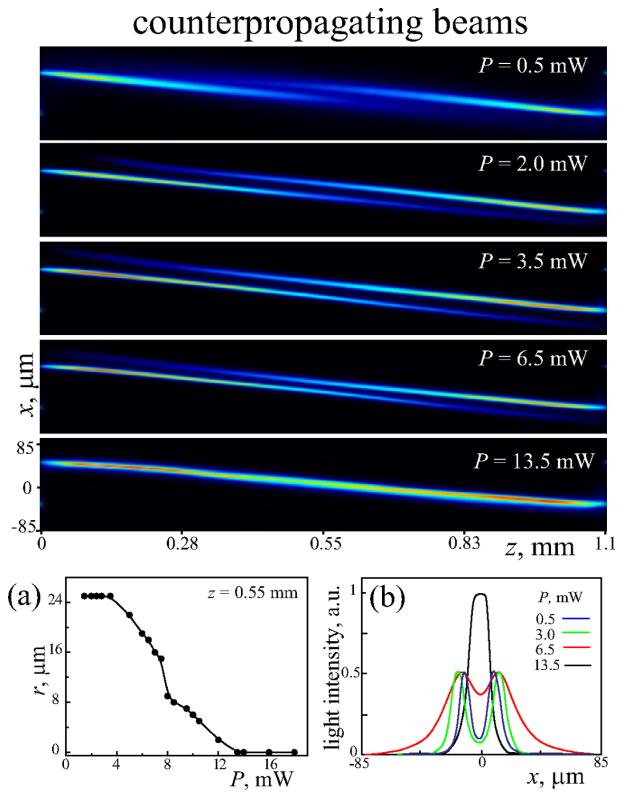}
\caption{Experimental results of the attraction between two CP solitons at different input powers. (a) Measured distance $r$ between solitons (at propagation distance $z=0.55$~mm) and corresponding (b) normalized intensity profiles for various input powers in each beam.}
\label{fig3}
\end{figure}

\section{Experimental results}

We employ the cell geometry sketched in Fig.~\ref{fig1}(a), with two parallel polycarbonate plates separated by a $100\mu m$ gap. The cell contains the 6CHBT liquid crystal~\cite{Baran, Dabrowski} which has negligible absorption and high nonlinearity with refractive indices $n_e$=1.6718 and $n_o$=1.5225 at room temperature. The polycarbonate plates are rubbed in the plane $(x,z)$ at the angle of $45^\circ$ with respect to the $z$ axis. Such prepared surfaces entail molecular orientation in a bulk analogous to that provided by pre-alignment via an external biasing field when the director is aligned along $z$. However, in contrast to the latter case where the director can rotate in a plane orthogonal to the plates, here the optical axis remains parallel to them.

The experimental setup is shown in Fig.~\ref{fig1}(c). Two identical beams generating spatial solitons are prepared using a system of mirrors and beam splitters. For CO solitons, two parallel input beams of extraordinary polarization (E-field along the $x$-axis) are formed using a standard Mach-Zehnder arrangement (beamsplitter BS1 and BS2, mirrors M1 and M2). In the case of CP solitons, the mirror M1 is removed and the beam transmitted by BS3 follows the route indicated by the dashed line. In addition, in the case of CO beams, the mirror M1 mounted on a piezoelectric transducer is made to vibrate (at frequency 1kHz) to induce a frequency shift on the reflected beam. Since the nonlinear response of NLC is slower than the mutual phase changes of two beams $\sim 10^{-3}$~s, such frequency detuning makes the interaction of beams effectively incoherent. With the half-wave plate we control the polarization state of the two identical and parallel Gaussian beams focused into the cell by a $10\times$ objective. The waist of each beam is $w_0=2\,\mu$m and the distance between them is $25\,\mu$m. The beam evolution inside the cell is monitored by collecting the light scattered above the cell with a $5\times$ microscope-objective and a high resolution CCD camera. The total beam power in all experiments was kept low (below 15~mW) in order to avoid thermooptical effects arising from light absorption and heating.

\begin{figure}
\centering\includegraphics[height=8cm]{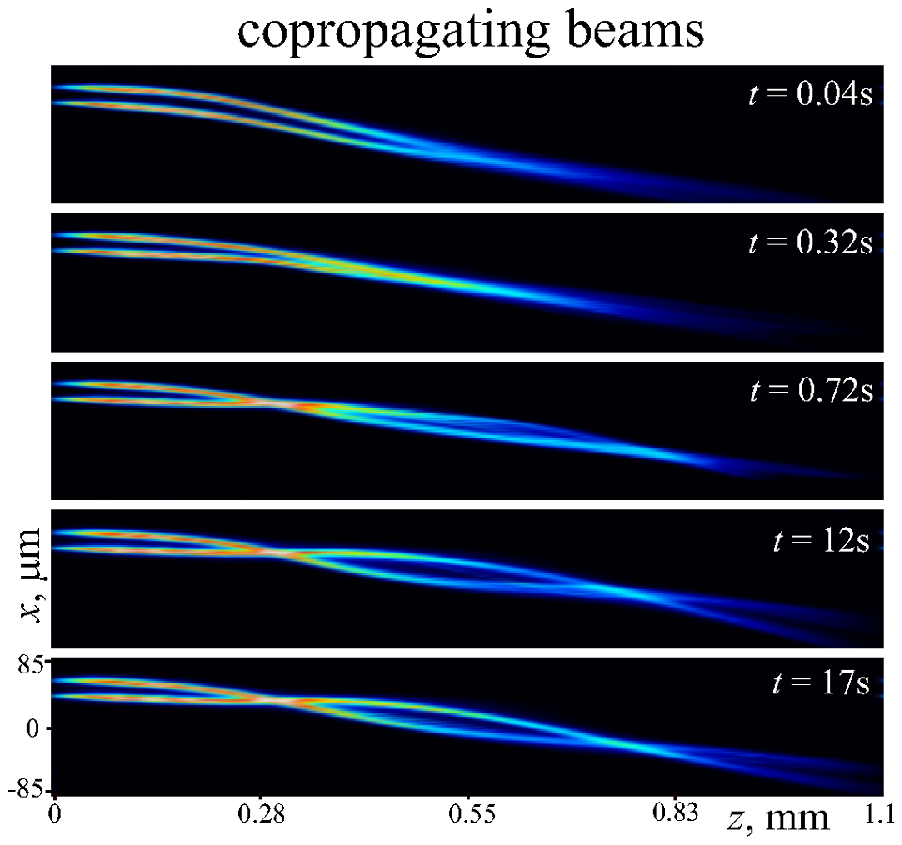}
\caption{{\color{blue} (Video 1)} Experimental results of the temporal evolution of two CO solitons shown in Fig.~\ref{fig2} for power 13.5~mW.}
\label{fig4}
\end{figure}

First study the dynamics of the formation and interaction of two CO parallel beams in NLC as a function of the excitation power, see Fig.~\ref{fig2}. As discussed earlier~\cite{Peccianti2002}, due to of the long-range character of nonlocality, the nematicons attract each other progressively more for increasing input power. In most cases this attraction independent of the relative phase of the solitons, even when they are coherent to each other. A sequence of images in Fig.~\ref{fig2} shows the stationary trajectories of two initially parallel beams for different input powers. For low input power (0.5~mW), the self-focusing is too weak to overcome diffraction, so the beams keep spreading as they propagate. By increasing the power to 2~mW, we achieve a stable propagation of solitons and their weak attraction. Due to high birefringence of the liquid crystal, both nematicons propagate at an angle with respect to the initial wave vector (directed along the $z$-axis). The measured walk-off of the Poynting vector is  $\approx 4.3^\circ$. For higher input power the attraction is so strong that it induces one (at $P\ge3$~mW) or multiple  ($P=13.5$~mW) intersections of the soliton trajectories.

Next we investigate the power-dependent dynamics of the CP solitons in the same cell. In this case the beams are launched from the opposite sides of the cell by focusing them with $10\times$ micro-objectives, see Figs.~\ref{fig1}(b,c). For better comparison with the CO case discussed above, we align the initial trajectories (without walk-off and without interaction) to be parallel to the optical axis $z$ and spatially separated by $25\,\mu$m. As in the case of CO nematicons, the waist of each CP beam is estimated to be $w_0=2\,\mu$m.

\begin{figure}
\centering\includegraphics[height=8cm]{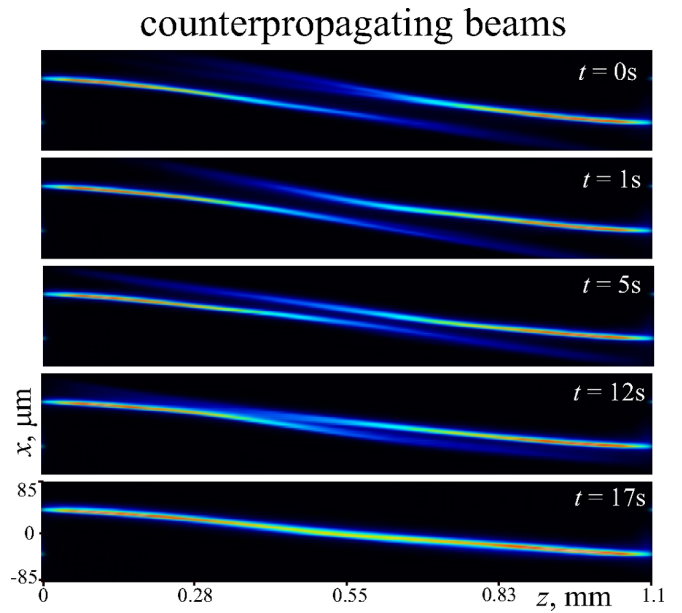}
\caption{{\color{blue} (Video 2)} Experimental results for the temporal evolution of two CP nematicons shown in Fig.~\ref{fig3} for power 13.5~mW.}
\label{fig5}
\end{figure}

Experimental results for CP beams are presented in Fig.~\ref{fig3}. As above, increasing the input power ($P\ge$2~mW) leads to self-focusing and the formation of two CP nematicons. However, by further increasing the power we observe a clear difference in the trajectories of the CP nematicons compared to the CO ones [cf. Fig.~\ref{fig2} and Fig.~\ref{fig3}]. In particular, we find that the CP trajectories remain almost parallel up to a power of 5~mW while the strong attraction in the CO case is already observed for about 2.5~mW. For higher powers, the trajectories approach each other until they eventually merge into a single soliton at 13.5~mW. In this regime both solitons propagate along the same trajectory forming the so-called vector soliton. This behavior is illustrated in Fig.~\ref{fig3}(b) where we plot the separation between trajectories of CP nematicons versus excitation power. It is clear that the trajectories start approaching each other only after the input power exceeds 5~mW. For higher powers, up to 18~mW, we do not observe changes in the soliton positions, neither associated with crossing, as in the CO case, nor because of spatiotemporal instabilities. Figure~\ref{fig3}(c) shows the intensity profiles for different input beam powers at the propagation distance of $z=0.55$~mm.

In order to understand the origin of the different dynamics of CO and CP nematicons, we have to consider the role of scattering losses, which progressively reduce the beam power and the size of the nonlinear effect. While the two CO beams evolve forward with equal individual powers at each $z$, the CP beams interact with unequal intensities and sizes, i.e. with unequal strengths in the transverse force pulling the two nematicons towards each other. The two initially parallel CP solitons, say A and B, launched at $z=0$ and $z=L$, respectively, tend to attract and shift the center of mass of the resulting dipole towards the more intense one, i.e. towards A for $z<L/2$ and towards B for $z>L/2$. This in-homogeneous force distribution results in a larger power requirement for the completion of the merging process as compared to the CO case.


\begin{figure}
\includegraphics[height=4.3cm]{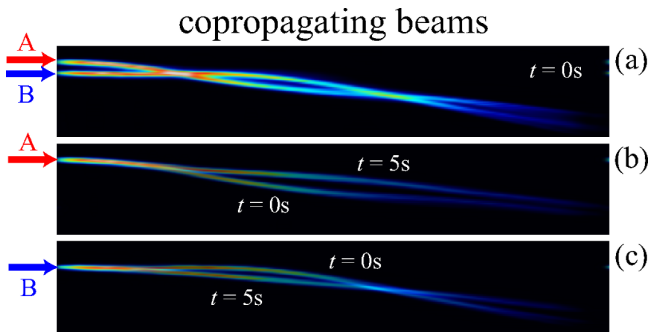}
\caption{(a) The bound CO nematicon pair at $t=0$ for the power of 13.5~mW and the initial separation of $25\,\mu$m (the same as in bottom frame of Fig.~\ref{fig2}). (b, c) Superimposed intensity profiles of a single beam in the pair (for $t=0$) and for $t=5$~s after the second beam has been blocked.}
\label{fig6}
\end{figure}

\begin{figure}
\includegraphics[height=4.3cm]{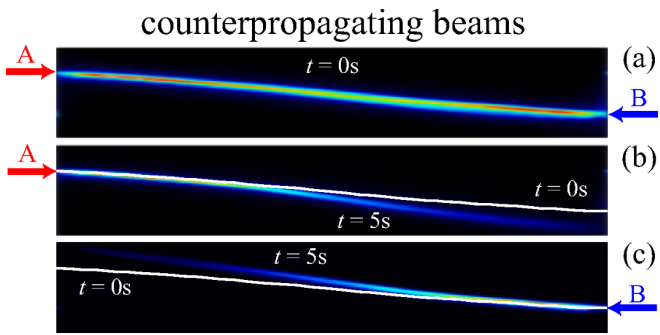}
\caption{The same as in Fig.~\ref{fig6} for CP nematicons from the bottom frame of Fig.~\ref{fig3}. The white line in (b) and (c) shows the position on the intensity maximum of the coupled pair in (a).}
\label{fig7}
\end{figure}

The asymmetry between CP nematicons has a profound effect on the temporal dynamics of interaction inducing significant differences with the CO case. Experimental sequence of images from Video 1, demonstrating the temporal evolution of two initially parallel CO beams, is presented in Fig.~\ref{fig4}. These images clearly reflect the inertia of the nonlinearity. After the light is switched on at $t=0$~s, the beams propagate linearly and diffract. It takes some time ($\sim0.3$~s) for the solitons to form. Only after additional time the soliton interaction becomes visible, leading to multiple intersections of their trajectories. In a sharp contrast, the temporal evolution of the CP beams is much slower as it requires 17~s for the solitons to merge and form a bound state, see Fig.~\ref{fig4} and Video 2.

We stress here that the differences between CO and CP nematicons are solely due to their interaction as all individual nematicons are equivalent in our experiments. To demonstrate this fact we study the relaxation dynamics of each beam after its neighbor is switched off at $t=0$. The results for CO (CP) nematicons are presented in Fig.~\ref{fig6} (Fig.~\ref{fig7}). To highlight the differences, in Figs.~\ref{fig6}(b,c) we superimpose two images of the same beam, with and without interaction, at $t=0$ and $t=5$~s respectively. Similarly, the two ``free'' CP nematicons in Figs.~\ref{fig7}(b,c) are compared to their common waveguide (white line). In both cases the figures demonstrate how the remaining soliton reconstructs its original individual trajectory after about 5~s. Importantly, this relaxation time is the same for both CO and CP cases because it is spent by a single nematicon to erase the changes in director orientation induced by both interacting beams. Thus, an order of magnitude difference in times necessary to build bound states in Figs.~\ref{fig4} and~\ref{fig5} should be considered as an essential signature of symmetric vs. asymmetric interactions of nematicons.

\vspace{-4mm}
\section{Conclusions}

We have investigated experimentally the generation, temporal dynamics, and power-dependence of interacting copropagating and counterpropagating spatial solitons in unbiased nematic liquid crystals outlining differences between these two cases. In particular, we have observed that the attraction of copropagating nematicons occurs at lower powers than for counterpropagating solitons, with an additional significant difference in the time response between the two cases.

This work was supported by the Australian Research Council.

\end{document}